\documentclass[aps, prl,amsmath,amssymb,twocolumn,floatfix,showpacs]{revtex4}
 
\usepackage{graphicx}
\usepackage{dcolumn}
\usepackage{bm}
\usepackage{epsfig} 

\newcommand{\bk}{{\bm k}} 
\newcommand{\bq}{{\bm q}} 
 
\newcommand{\bx}{{\bm x}}

\graphicspath{{./figs/}}

\begin{document} 
 
\preprint{For submission to {\it Physical Review Letters}} 
 
\title{Three-dimensional Phase Field Quasicrystals}
 
\author{P.~Subramanian$^1$, A.J.~Archer$^2$, E.~Knobloch$^3$ and A.M.~Rucklidge,$^1$}
\affiliation{$^1$Department of Applied Mathematics, University of Leeds, Leeds LS2 9JT, UK\\
$^2$Department of Mathematical Sciences, Loughborough University, Loughborough LE11 3TU, UK\\
$^3$Department of Physics, University of California at Berkeley, Berkeley, CA 94720, USA}

\date{\today}

\begin{abstract} 
We investigate the formation and stability of icosahedral quasicrytalline
structures using a dynamic phase field crystal model. Nonlinear interactions
between density waves at two length scales stabilize three-dimensional
quasicrystals. We determine the phase diagram and parameter values required
for the quasicrystal to be the global minimum free energy state. We
demonstrate that traits that promote the formation of two-dimensional
quasicrystals are extant in three dimensions, and highlight the
characteristics required for 3D soft matter quasicrystal formation.

\end{abstract} 
 

 
\maketitle 


Regular crystals form ordered arrangements of atoms or molecules with rotation and
translation symmetries, and possess discrete X-ray diffraction patterns, or equivalently, discrete spatial Fourier spectra. In contrast, quasicrystals (QCs) lack the translational
symmetries of regular crystals, yet also display discrete spatial Fourier spectra. QCs made
from metal alloys were discovered in 1982~\citep{ShectmanPRL1984} and attracted the
Nobel prize for chemistry in 2011. QCs can be quasiperiodic in all three dimensions
(usually with icosahedral symmetry), or can be quasiperiodic in two (or one) directions
while being periodic in one (or two). The vast majority of the QCs discovered so far
are metallic alloys (e.g., Al/Mn or Cd/Ca). However, QCs have recently been found in
nanoparticles~\citep{Talapin2009}, mesoporous silica~\citep{Xiao2012}, and
soft-matter~\citep{Dotera2011} systems. The latter include micellar
melts~\citep{Zeng2004,Fischer2011} formed, e.g., from linear, dendrimer or star block
copolymers.

In recent years, model systems in two dimensions (2D) have been studied in order to understand soft-matter QC formation and stability~\citep{Barkan2011,Archer2013,Dotera2014,Barkan2014,Archer2015}. The ingredients for 2D QC formation are, firstly, a propensity towards periodic density modulations with two characteristic
wave numbers $k_1$ and $k_2$. The ratio $k_2/k_1$ must be close to certain special values;
e.g., for dodecagonal QCs the value is $2\cos\frac{\pi}{12}$. Secondly, strong reinforcing
(i.e., resonant) nonlinear interactions between these two characteristic density waves
are required~\cite{Lifshitz1997,RucklidgePRL2012,Lifshitz2015}.
Here we demonstrate that analogous traits promote the formation of
three-dimensional (3D) soft matter QCs with icosahedral symmetry,
namely resonant nonlinear interactions involving two length scales that are
within a factor of two of each other.
Nonlinear resonant interactions between density waves at a single wavelength
are important in stabilizing simple crystal structures, such as body-centered cubic (bcc)
crystals~\citep{Alexander1978} and with the right coupling, even QCs~\citep{BakPRL1985}.

We consider a 3D~phase field crystal (PFC) model that generates modulations with
two length scales. The PFC model predicts the density
distribution of the matter forming a solid or a liquid on the microscopic length scale
of the constituent atoms or molecules, and takes the form of a theory for a dimensionless scalar
field $U(\bx,t)$ that specifies the density deviation from its average value at position
$\bx$ at time $t$ \cite{Emmerich2012}.  This {model} consists of a nonlinear partial differential equation (PDE) with conserved dynamics, describing the time evolution of $U$ over diffusive time scales \cite{Emmerich2012}. 
Our PFC model includes all the resonant interactions that occur in the case of icosahedral symmetry.
This not only extends previous work to three
dimensions, but also allows for independent control over the growth rates of waves with
the two wavelengths, and shows that, just as for 2D~QCs, resonant interactions between
the two wavelengths do stabilize 3D~QCs.


Our PFC model is based on writing the free energy~$\mathcal{F}$ as
 \begin{equation}
 {\mathcal F}\left[U\right]=\int \Big[
           -\frac{1}{2}U{\mathcal L}U - \frac{Q}{3}U^3 + \frac{1}{4}U^4
           \Big]d\bx\,,
 \label{eq:fe}
 \end{equation}
where the operator ${\mathcal L}$ and parameter $Q$ are defined below. The evolution equation for $U$ 
follows conserved dynamics and can be
obtained from the free energy as
 \begin{equation}
 \frac{\partial U}{\partial t}=\nabla^2\left(\frac{\delta \mathcal{F}[U]}{\delta U}\right)=-\nabla^2\left(\mathcal{L}U+QU^2-U^3\right).
 \label{eq:evol}
 \end{equation}
This evolution equation describes a linearly unstable system that is stabilized nonlinearly by the
cubic term. The relative importance of second order resonant interactions can be varied by setting 
the value of~$Q$. The average value of $U$ is conserved, so $\bar U$ is effectively a parameter of
the system. Without loss of generality we choose $\bar U=0$, since other values can be accommodated by altering $\mathcal{L}$
and~$Q$.

   \begin{figure}
 {\includegraphics[width=0.75\hsize]{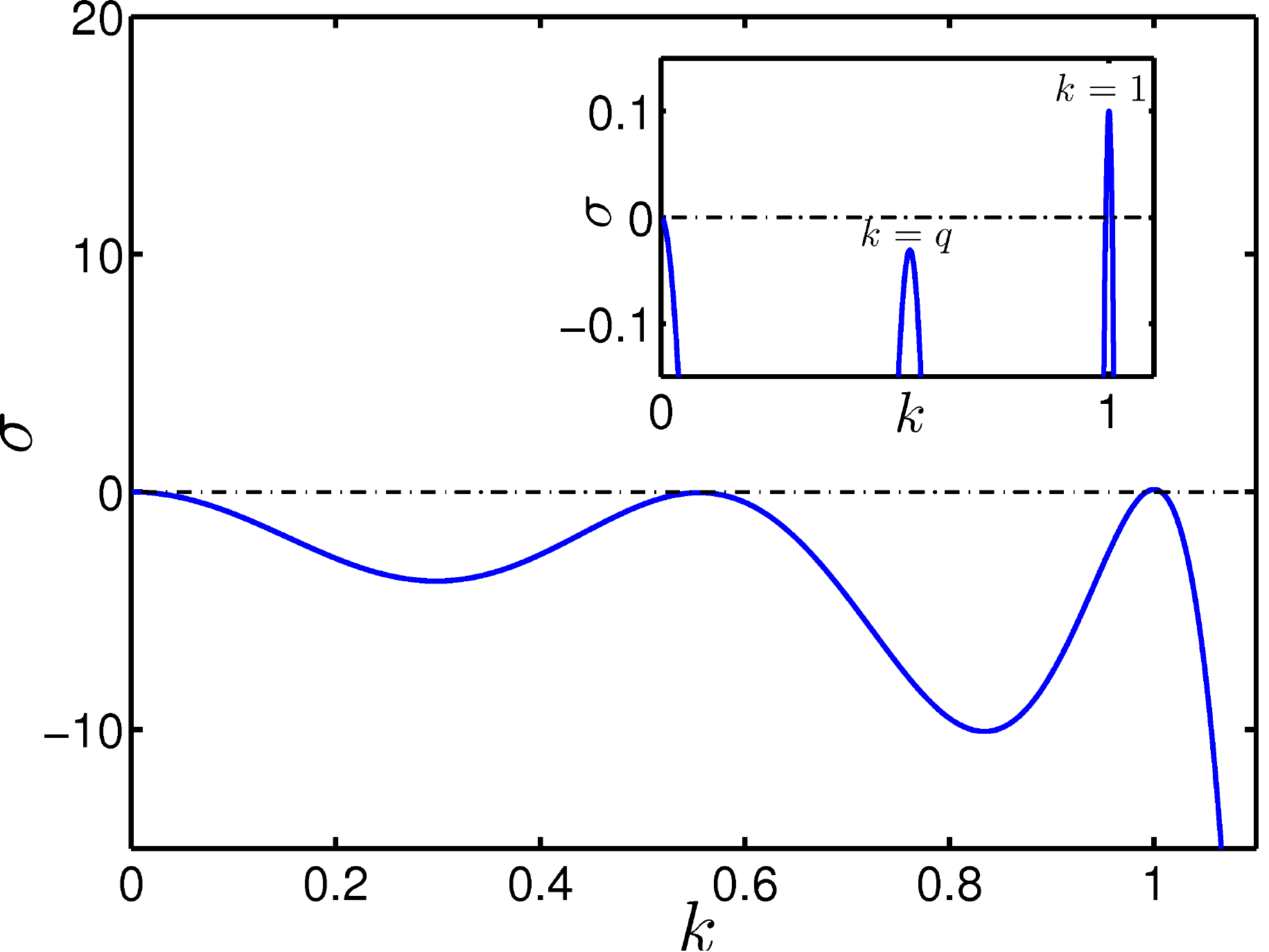}\hfil}
 \vspace{-2ex}
 \caption{\label{fig:disprel} Growth rate $\sigma(k)$ as a function of the wave number~$k$ 
for the linear operator~$\mathcal{L}$ in Eq.~(\ref{eq:evol}), as defined in Eq.~(\ref{eq:disprel}), with
parameters $\sigma_0=-100$, $q=1/\tau=0.6180$, $\mu=0.1$
and $\nu=-0.1$. The growth rates at $k=1$ and $k=q$ are $\mu$ and $q^2\nu$, as in the inset.}
 \end{figure}

The model is based on the original PFC model of \citet{ElderPRL2002}, which allowed
linear instability at a single length scale, and was stabilised by a cubic term. Subsequently, \citet{AchimPRL2014} used ideas based on the Lifshitz--Petrich model \citep{Lifshitz1997} to extend the problem to
include two length scales. However, the growth rates of the two length scales in their models were
constrained to be in a fixed ratio. In our model, we choose the linear operator $\mathcal{L}$ (based on the one introduced by \citet{RucklidgePRL2012}), to allow marginal instability at two wave numbers $k=1$ and $k=q$, with the growth rates of the two length scales determined by two {independent} parameters $\mu$ and $\nu$, respectively. The resulting growth rate $\sigma(k)$ of a mode with wave number~$k$ is given by a tenth-order polynomial:
 \begin{equation}
 \sigma(k)=\frac{k^4[\mu A(k)+\nu B(k)]}{q^4(1-q^2)^3}+\frac{\sigma_0 k^2}{q^4}(1-k^2)^2\,(q^2-k^2)^2\,,
 \label{eq:disprel}
 \end{equation} 
where $A(k)=[k^2(q^2-3)-2q^2+4](q^2-k^2)^2 q^4$ and $B(k)=[k^2(3q^2-1)+2q^2-4q^4](1-k^2)^2$.
Figure~\ref{fig:disprel} shows a typical $\sigma(k)$, with $k=0$ neutrally {stable} and $k=q,1$
weakly stable and unstable, respectively. The operator~$\mathcal{L}$ is obtained
from Eq.~(\ref{eq:disprel}) by replacing $k^2$ by $-\nabla^2$.
 
The PFC model defined in Eq.~(\ref{eq:evol}) can be used to explore the effect of resonant triadic
interactions on the resulting final structure. We encourage structures with icosahedral 
symmetry by setting the value of the wave number ratio $q=1/\tau$, where $\tau=2\cos\frac{\pi}{5}=1.6180$ is the golden ratio. The other parameters are $\sigma_0$, $\mu$, $\nu$ and~$Q$. In the rest of this paper, we set $\sigma_0=-100$ to ensure that the maxima in growth rate are sharp, and $Q=1$, a value that is large enough for effective nonlinear interactions while still being amenable to weakly nonlinear analysis. We analyze the system in the remaining $2$-parameter space, varying $\mu$ and $\nu$ simultaneously.

\begin{figure*}[]
 \hbox to \hsize{\hfil\hbox to 0.225\hsize{(a)\hfil}\hfil
                      \hbox to 0.225\hsize{(b)\hfil}\hfil
                      \hbox to 0.225\hsize{(c)\hfil}\hfil
                      \hbox to 0.225\hsize{(d)\hfil}\hfil}
 \vspace{-3ex}
 \hbox to \hsize{\hfil
 \includegraphics[width=0.205\hsize]{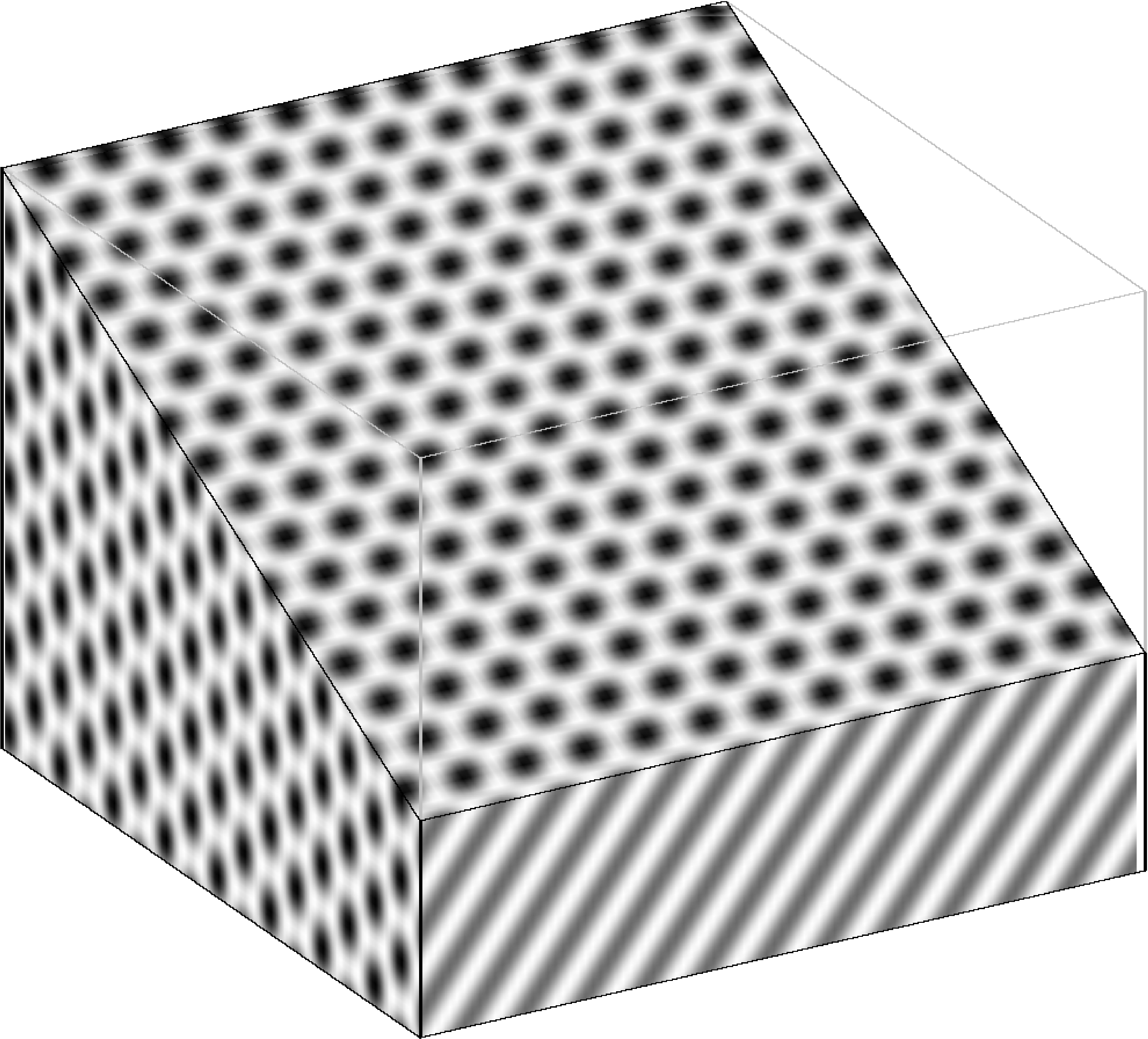}\hfil
 \includegraphics[width=0.205\hsize]{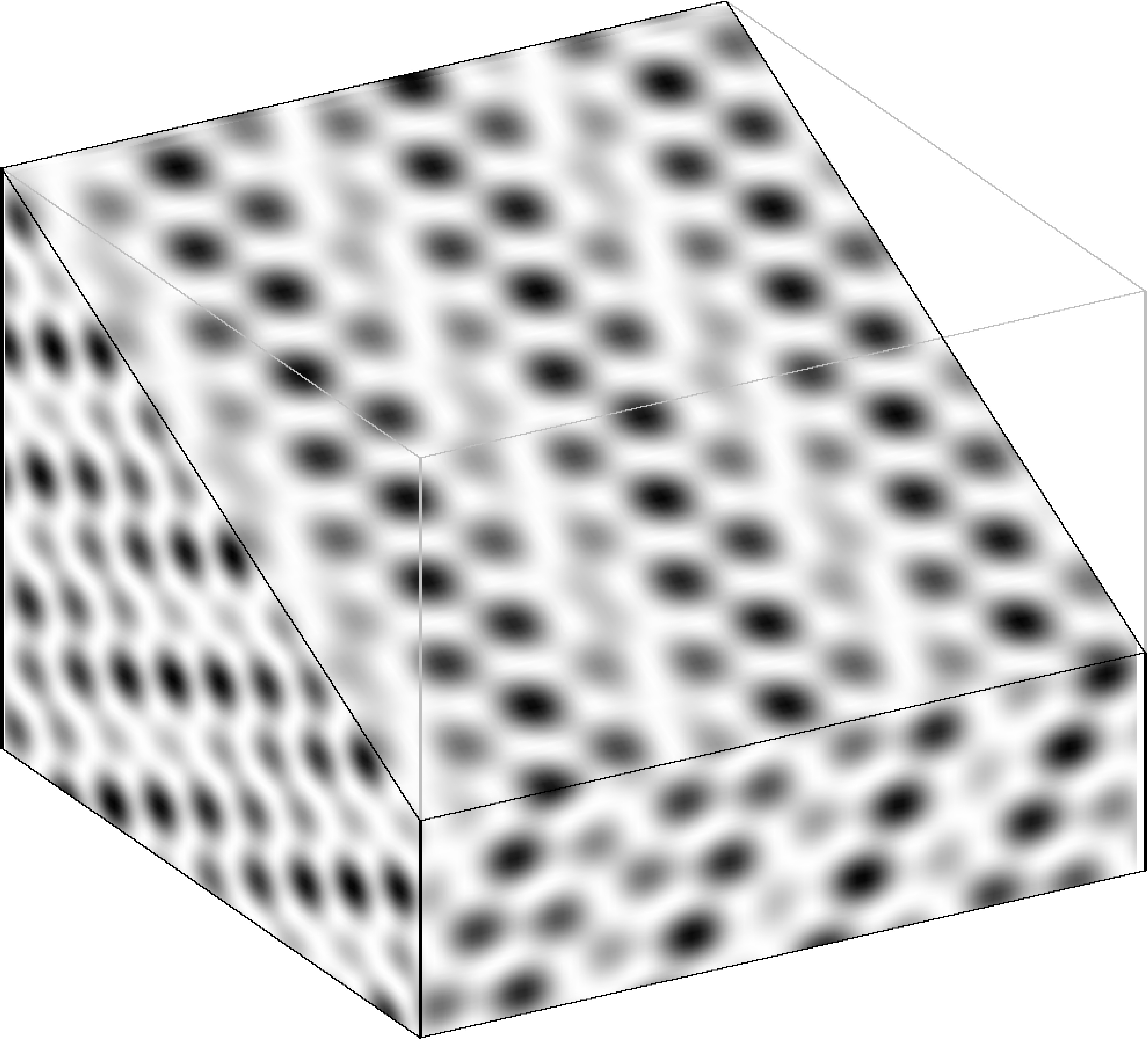}\hfil
 \includegraphics[width=0.205\hsize]{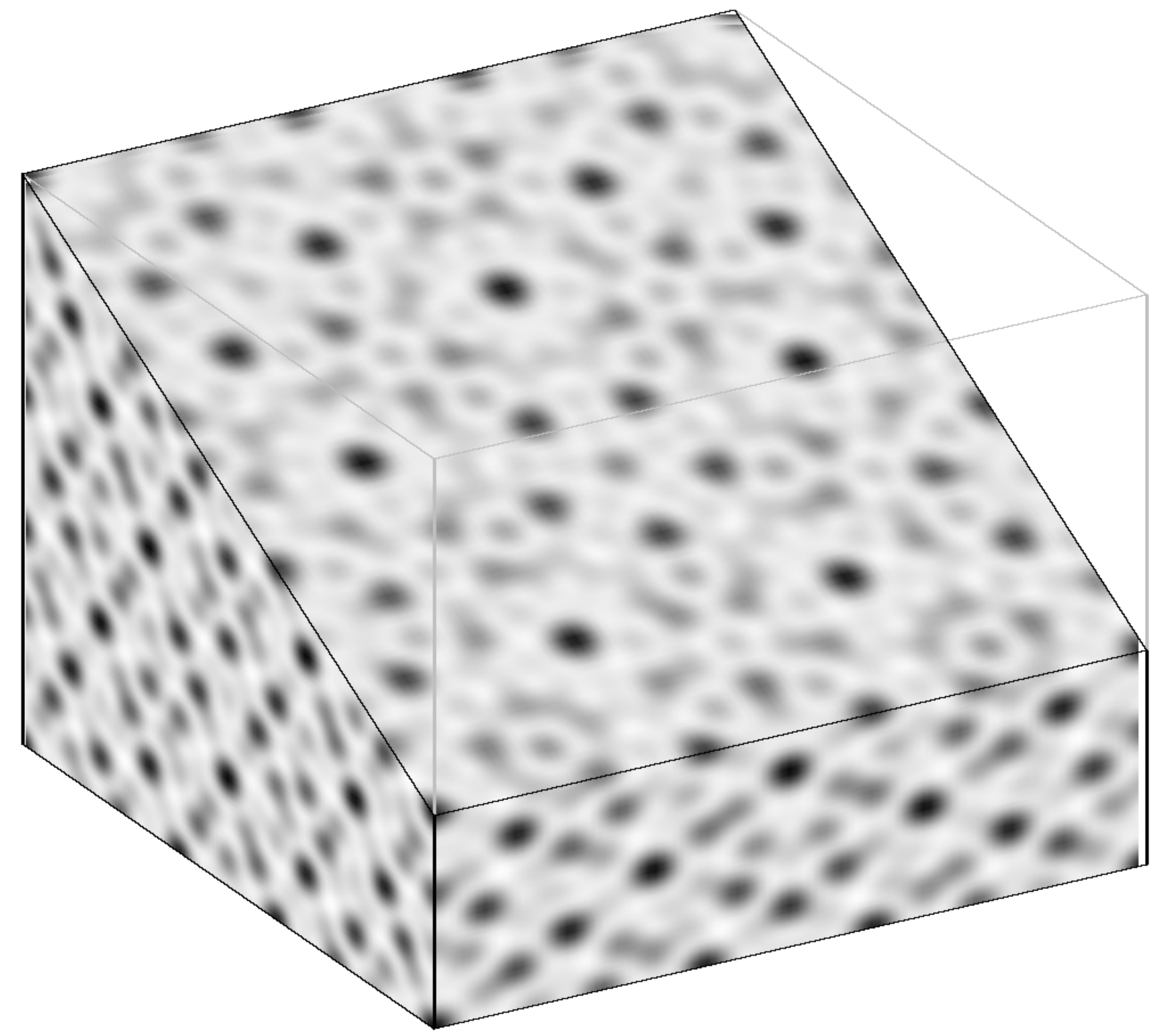}\hfil
 \includegraphics[width=0.182\hsize] {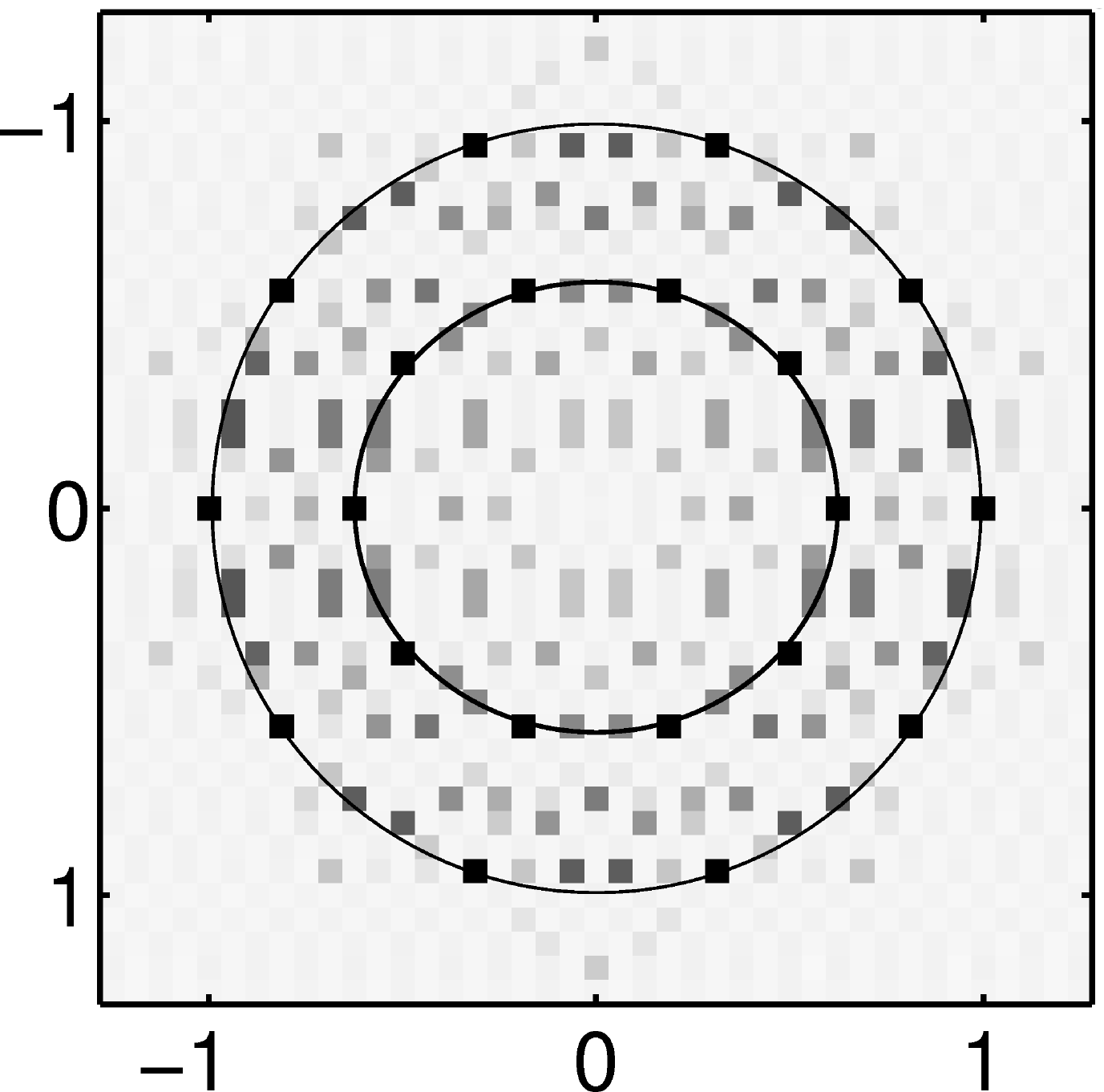}\hfil}
\vspace{-2ex}
 \caption{\label{fig:simulations}
 (a)~Hexagonal columnar phase with wave number~$q$ ($q$-hex) at $(\mu,\nu)=(0.082,0.056)$.
 (b)~Body centered cubic crystal with wave number~$1$ ($1$-bcc) at $(\mu,\nu)=(-0.1,0)$.
 (c)~Icosahedral quasicrystal (QC) at $(\mu,\nu)=(-0.071,-0.071)$.
Each box has had a slice cut away, chosen to reveal the 5-fold rotation symmetry in~(c). See \citep{Suppl1} for more details on the quasicrystalline structure.
 (d)~Diffraction pattern taken in a plane normal to the vector $(\tau,-1,0)$ in Fourier space. The circles of radii $1$ and $q$ are indicated. The 10-fold rotation symmetry of the diffraction pattern is indicated by the 10~peaks observed on each circle.}
 \end{figure*}

Three-dimensional direct numerical simulations of the PDE~(\ref{eq:evol}) were carried out in a
periodic cubic domain of size $(16\times2\pi)^3$, corresponding to $16$~of the shorter of the two
wavelengths. This choice is guided by the fact that domains that are twice a Fibonacci number (in this case~8) allow our periodic solutions to approximate true quasicrystals well. We used $192$ Fourier modes (using FFTW \citep{FFTW05}) in each direction and employed second-order exponential time
differencing (ETD2) \citep{CoxJCP2002}. Simulations were carried out for $32$~combinations of
$\mu$ and $\nu$ lying on a circle of radius~$0.1$ in angular steps of $\Delta\theta=11.25^{\circ}$.
The
simulations were started from an initial condition consisting of smoothed random values with an
amplitude of $\mathcal{O}(10^{-3})$ for each Fourier mode, and evolved to an asymptotic state. In
cases where the {solution did} not decay to the zero flat state (corresponding to the uniform liquid), 
qualitatively {distinct} asymptotic
states {were} found. These {include} hexagonal columnar crystals (hex), body-centered cubic crystals (bcc)
at each of the two wavelengths, in addition to a three-dimensional icosahedral
quasicrystal. Examples of $q$-hexagons, $1$-bcc and the icosahedral quasicrystal are shown in
Figs.~\ref{fig:simulations}(a)--(c). Figure~\ref{fig:simulations}(d) shows a diffraction pattern with 10-fold symmetry for the quasicrystal.

\begin{figure}[b]
 \hbox to \hsize{\hfil\hbox to 0.45\hsize{(a)\hfil}\hfil
                      \hbox to 0.45\hsize{(b)\hfil}\hfil}
 \vspace{-3ex}
 \hbox to \hsize{\hfil
 \includegraphics[width=0.45\hsize]{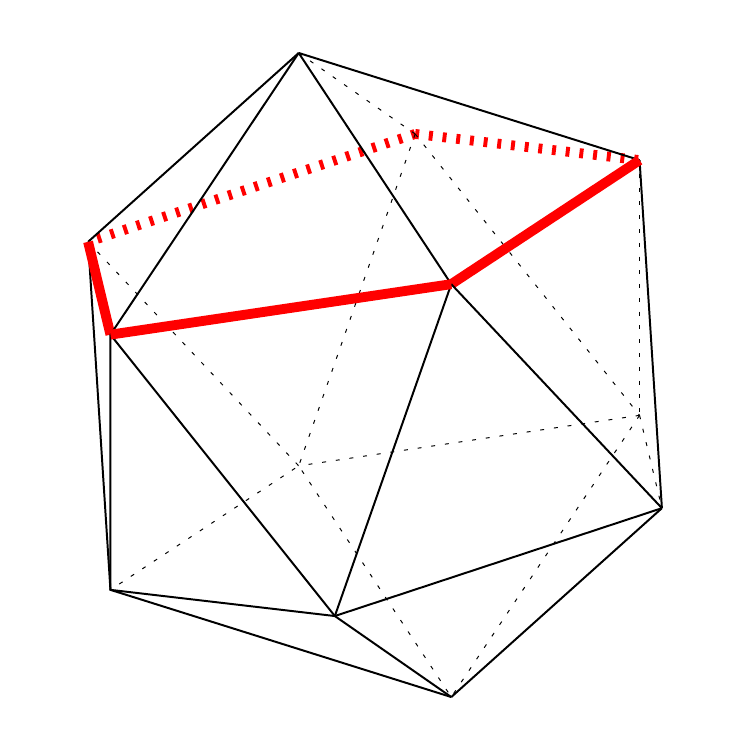}\hfil
 \includegraphics[width=0.45\hsize]{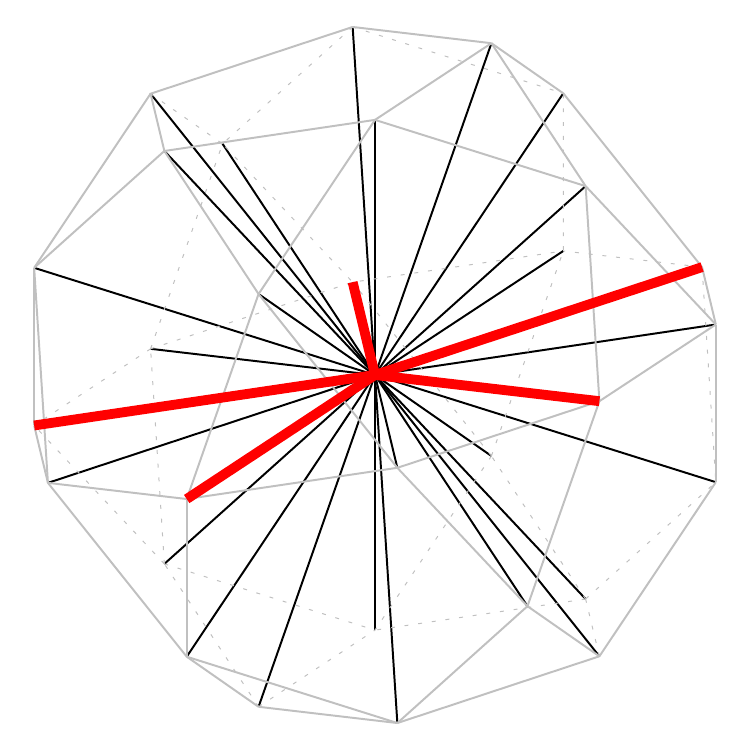}\hfil}
\vspace{-2ex}
 \caption{\label{fig:icosahedra} (a)~Icosahedron, with five edges
indicated with thick red lines (color online) such that the five edge vectors
add up to zero. (b)~With the 30 edge vectors moved to the origin
(the same five vectors are indicated), the resulting
geometric figure, made by connecting the ends of these 30 vectors, is an
icosidodecahedron, with 20 triangular faces and 12 pentagonal faces. Edges that
are behind each figure are indicated with dotted lines.}
 \end{figure}

The stability of quasicrystals is promoted by nonlinear wave interactions, of three or more waves.
In Ref.~\cite{BakPRL1985}, it is pointed out that density perturbation waves (at one length scale) of
the form $e^{i\bk\cdot\bx}$ with wavevectors chosen to be the 30~edge vectors of an icosahedron can
take advantage of three-wave interactions (from the triangular faces) and of five-wave interactions
(from the pentagons surrounding five triangular faces, see Fig.~\ref{fig:icosahedra}(a)) to lower
the free energy and so encourage the formation of icosahedral quasicrystals. This results in having
density waves involving 30~wavevectors, see Fig.~\ref{fig:icosahedra} and Table~\ref{tab:vec}.

With two length scales in the golden ratio~$\tau$, an alternative mechanism for reinforcing
icosahedral symmetry is possible using only three-wave interactions. Taking
five edge vectors of a pentagon adding up to zero, for example, $\bk_{16}+\bk_7+\bk_{15}+\bk_2+\bk_{25}=0$ 
(see Table~\ref{tab:vec}),
we use the fact that $\bk_7+\bk_{15}=\bq_2$ and $\bk_2+\bk_{25}=\bq_4$ {to identify} a three-wave
interaction between $\bq_2$, $\bq_4$ and $\bk_{16}$ since these sum to zero. Many other three-wave 
interactions are possible.

 \begin{table}[b]
\centering{\begin{tabular}{c | c || c | c || c | c} $j$ & $\bk_j$ & $j$ & $\bk_j$ & $j$ & $\bk_j$\\ \hline                                                                                                                                                                \\[-8pt]
\mathstrut{1} & $(1, 0, 0)$                     &  {6} & $\frac{1}{2} (1,\tau-1,\tau) $    & {11} & $\frac{1}{2}(\tau-1, \tau,-1)$ \\[3pt]
\mathstrut{2} & $\frac{1}{2}(\tau,  1, \tau-1)$ &  {7} & $\frac{1}{2} (1,\tau-1,-\tau) $    & {12} & $\frac{1}{2}(\tau-1,-\tau,-1)$ \\ [3pt]
\mathstrut{3} & $\frac{1}{2}(\tau, 1,1-\tau)$  &  {8} & $\frac{1}{2}(1,1-\tau,-\tau)$ & {13} & $\frac{1}{2}(\tau-1,-\tau, 1)$ \\[3pt] 
\mathstrut{4} & $\frac{1}{2}(\tau, -1,1-\tau)$ &  {9} & $\frac{1}{2}(1,1-\tau, \tau)$ & {14} & $(0, 1, 0)$ \\[3pt] 
\mathstrut{5} & $\frac{1}{2}(\tau, -1, \tau-1)$ & {10} & $\frac{1}{2}(\tau-1,\tau, 1)$  & {15} & $(0, 0, 1)$ \\[3pt] 
\hline
 \end{tabular}}
 \caption{Indexed table of edge vectors $\bk_1,\dots,\bk_{15}$ of an icosahedron with edges of 
length~1, following Ref.~\cite{Levine1986}.
 The remaining 15 are the negatives: $\bk_{j+15}=-\bk_j$. 
 The 30 vectors on the other sphere, of radius $q=\frac{1}{\tau}$, 
 are obtained by setting
 $\bq_j=\bk_j/\tau$, $j=1,\dots,30$.}
 \label{tab:vec}
 \end{table}
We can now analyse the icosahedral quasicrystals shown in Fig.~\ref{fig:simulations}(c). At small amplitudes, $U$ can be rescaled in terms of a small parameter $\epsilon$ as $U=\epsilon U_1$. Substituting this into the expression for the free energy and requiring that the three terms contribute at the same order implies a scaling $Q=\epsilon Q_1$ and $\mathcal{L}U=O(\epsilon^3)$. The scaling of the linear operator can be arranged by requiring that $U_1$ is a combination of Fourier modes with wave numbers $k=1$ and $k=q$ and that the parameters $\mu$ and $\nu$, which govern the linear growth rates of these two wave numbers, scale as $\mathcal{O}(\epsilon^2)$. Upon substituting these expressions into Eq.~(\ref{eq:evol}), we observe that the time evolution occurs on slow time scales, of order $\mathcal{O}(\epsilon^{-2})$.
 
For icosahedral quasicrystals, we use the vectors from Table~\ref{tab:vec} and expand
$U_1$ as
\begin{equation}
U_1=\sum_{j=1}^{15} z_j\, e^{i {\bk_j} \cdot {\bx}} + 
    \sum_{j=1}^{15} w_j\, e^{i {\bq_j} \cdot {\bx}} + c.c.,
\end{equation}
where $c.c.$ refers to the complex conjugate, the amplitudes $z_j$ and $w_j$ are
functions of time and describe the evolution of modes with wave numbers 1 and $q$, respectively.

Substituting this expression for $U_1$ into Eq.~(\ref{eq:fe}), we can write the rescaled volume specific
free energy $f=\mathcal{F}/(V\epsilon^4)$ as
 \begin{align} f=-\mu z_1 \bar{z}_1 &- 4Q
\big( w_{10} z_4 - w_{11} z_5 - w_{12} z_2 - w_{13} z_3 \nonumber \\
 &\phantom{4Q(}{}- w_3 w_5 - w_2 w_4 - z_6 z_8 - z_7 z_9  \big) \bar{z}_1 \nonumber \\
&{}-\mu\sum_{j=2}^{15} |z_j|^2 -\nu\sum_{j=1}^{15} |w_j|^2 \nonumber \\
&{}-Q(152~\textrm{other~cubic~terms})\nonumber \\
&{}-(1305~\textrm{quartic~terms}),
\label{eq:smallf}
\end{align}
where we have written the contributions involving $\bar{z}_1$ explicitly up to cubic order. All other
contributions are of similar structure. Nonlinear terms at every order $n$ contain combinations of $n$
vectors that sum to zero. The evolution on the slow
time scale of the amplitudes of the components of $U_1$ {is thus governed by the equations}
 \begin{equation}
 \dot{z}_j=-\frac{\partial f}{\partial \bar{z}_j}
 \qquad\mbox{and}\qquad
 \dot{w}_j=-q^2 \frac{\partial f}{\partial \bar{w}_j}.
 \label{eq:ampeqns}
\end{equation}
These evolution equations are the projection of the PDE (\ref{eq:evol}) onto the 60 Fourier modes.

It is straightforward to find subsets of non-zero amplitudes that give equilibrium solutions that 
describe simple structures, such as lamellae (lam), rhombi, hexagonal (hex) columnar crystals, and
simple cubic crystals, at each length scale. More complex structures typically involve both length
scales; these include 2D planar quasicrystals (possibly periodic in the third direction), and 3D
quasicrystals with icosahedral or five-fold symmetry. 

Within each class of solutions, we write down amplitude equations restricted to that class and solve
the resulting coupled algebraic equations to obtain equilibrium solutions using the Bertini numerical
algebraic geometry software package~\citep{Bertini}. Using expression (\ref{eq:smallf}), we calculate
the minimum free energy~$f$ associated with each class of solutions. By minimizing this over all
classes of solutions at a given combination of $\mu$ and~$\nu$, we calculate the globally stable
solution. Since we found body-centered cubic (bcc) crystals in Fig.~\ref{fig:simulations}(b), and since
these cannot be represented in terms of the icosahedral basis vectors, we compute their free energy as
a separate calculation, choosing a different set of basis vectors~\citep{CallahanNL1997}.
 
  \begin{figure}
 {\includegraphics[width=0.65\hsize]{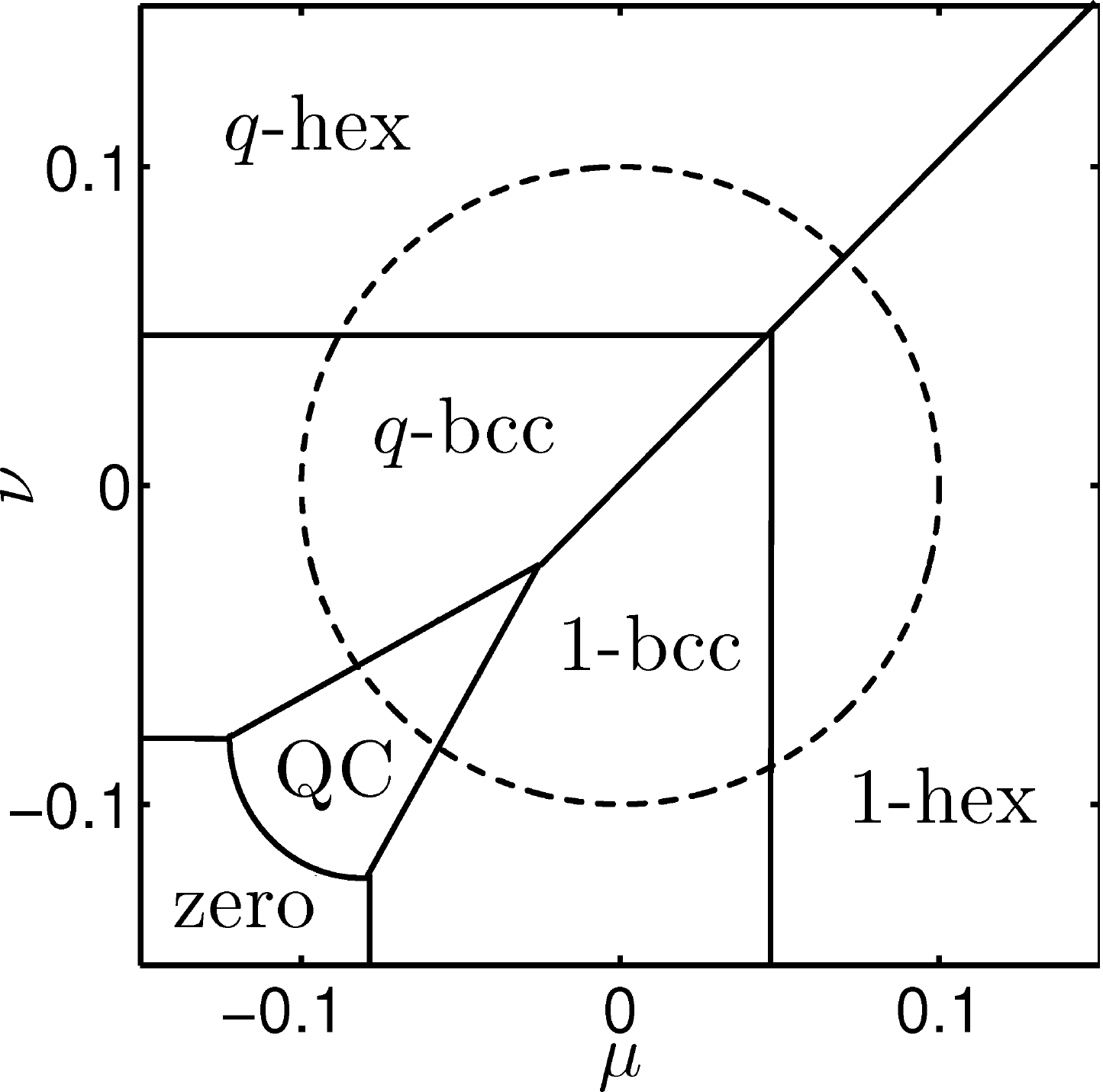}\hfil}
\vspace{-2ex}
 \caption{\label{fig:minfe}
 Structures with minimal specific free energy~$f$ over a range of
parameters $\mu$ and~$\nu$, computed as equilibria of the amplitude equations~(\ref{eq:ampeqns}).
PDE calculations are performed on the dashed circle around the origin with radius $0.1$. 
The region in the third quadrant labeled `zero' indicates that the 
trivial state $U=0$ is globally stable.}
 \end{figure}

Figure~\ref{fig:minfe} shows regions in the $(\mu,\nu)$ plane, identifying the globally stable solution
in each region. Body-centered cubic and hexagonal columnar crystals are observed at both
wavelengths independently, and their regions of global stability are symmetric with respect to the
$\mu=\nu$ line. At larger values of $\mu$ and $\nu$, the regions of $1$-hex and $q$-hex are {bounded, likewise
symmetrically}, by lamellar patterns $1$-lam and $q$-lam, above the lines $\mu=1.91$ and
$\nu=1.91$, respectively. The symmetry with respect to the $\mu=\nu$ line is a consequence of the
particular structure of the model. The zero region in the third quadrant indicates that the trivial
state $U=0$ is globally stable. Three-dimensional quasicrystals are observed as the global minimum for
cases when both the linear growth rates $\mu$ and $\nu$ are negative.
This region of global stability of QCs vanishes when~$Q=0$.

The local (linear) stability of the equilibria is obtained by linearizing the amplitude equations~(\ref{eq:ampeqns}).
The regions of local stability extend beyond the lines {demarcating the boundaries} of the regions of
global stability, and many locally stable structures can coexist at given parameter combinations.

\begin{figure}
 {\includegraphics[width=0.91\hsize]{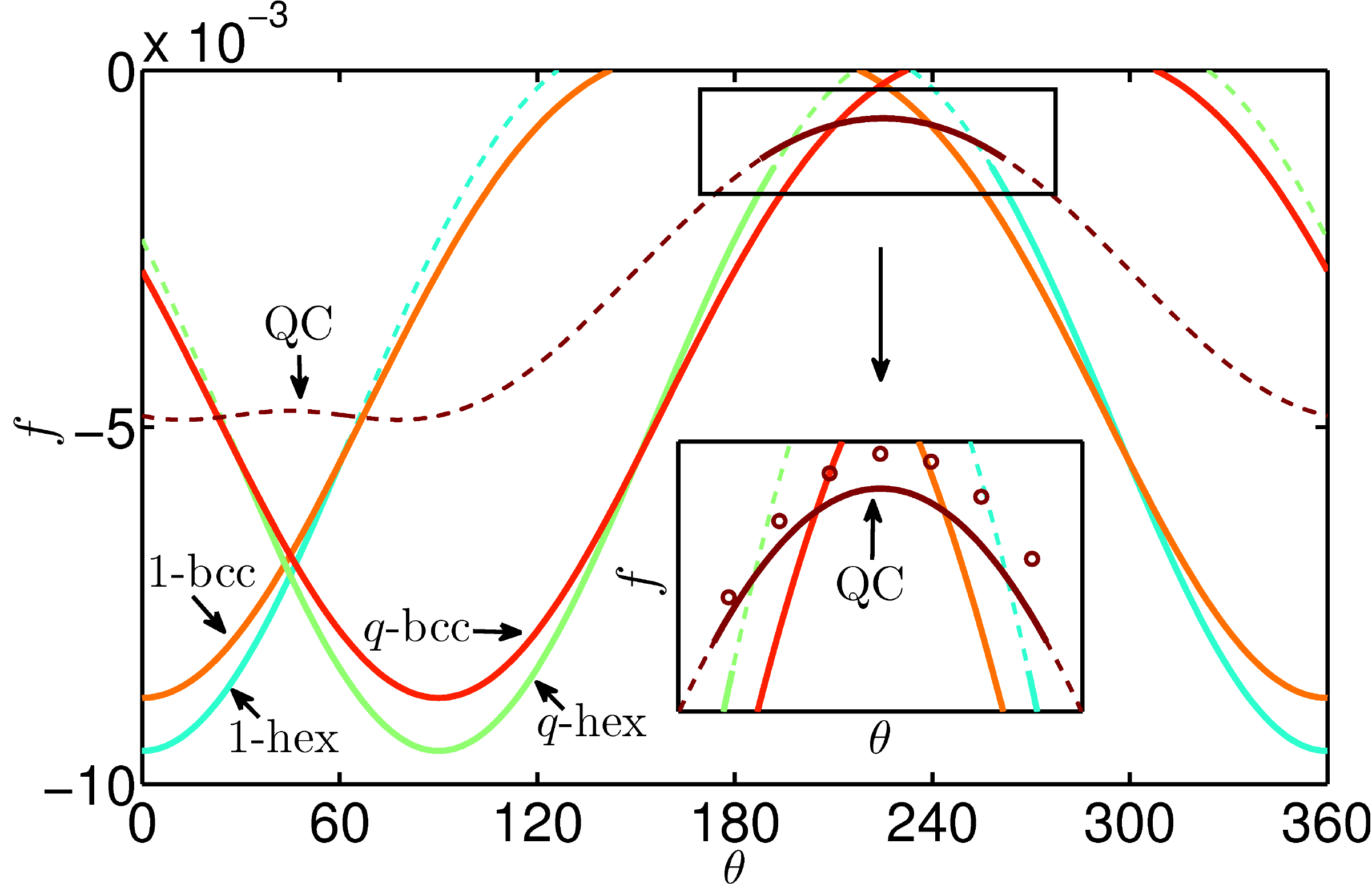}\hfil}
 \caption{\label{fig:fevarstab}
(Color online) Variation of specific free energy~$f$ with angle $\theta$ on a circle in the $(\mu,\nu)$ plane of radius $0.1$. Lines track the variation of free energy~$f$ of the labeled structures, solid where these are locally stable, dashed where they are locally unstable. We do not make this distinction for the bcc crystals as these use a different set of basis vectors and so their linear stability cannot be compared directly with that of quasicrystals. The zero state, $f=0$, corresponds to the uniform liquid. Hollow circles in the inset show the free energies of locally stable quasicrystalline asymptotic steady states from PDE calculations starting from an initial condition of the imprinted quasicrystal.}
 \end{figure}

Figure~\ref{fig:fevarstab} shows the variation of the specific free energy~$f$ in Eq.~(\ref{eq:smallf}) 
around the dashed circle shown in Fig.~\ref{fig:minfe}. We focus on negative free energies only (i.e., 
states with energy lower than the uniform density liquid state), and from the figure we
can read off the parameter range where each structure emerges as the global minimum. 
In spite of the large number of
three-wave interactions in the icosahedral structure, 3D quasicrystals emerge as globally stable
states only over a limited range of angles ($213.53^{\circ}\leq \theta \leq 236.47^{\circ}$). 
In the range of parameters investigated here, 2D~planar quasicrystals (not shown) are never globally
stable.

Hollow circles in the inset in Fig.~\ref{fig:fevarstab} show the free energies of locally stable
quasicrystalline steady states of the PDE~(\ref{eq:evol}), started from an initial condition with the
quasicrystal imprinted. 
The fact that the solid line for the quasicrystalline free energy (from the
small~$\epsilon$ asymptotics) is close to the hollow circles (from the PDE), both with respect to the
value of the free energy and the range of linear stability, supports the validity of the asymptotics,
{despite the mathematical subtleties associated with QCs, identified} in \citep{Rucklidge2003}, and 
partly resolved in \citep{Iooss2010}.


The parameters~$Q$ and $\sigma_0$ were chosen so as to allow good agreement
between minima of the free energy~(\ref{eq:fe}) and {its weakly nonlinear
approximation derived in} Eq.~(\ref{eq:smallf}). This agreement, and the 
prediction from the asymptotics that 
the region where QCs are globally stable vanishes
when $Q=0$, confirms that the contribution to the free energy from three-wave
interactions is crucial in stabilizing 3D icosahedral~QCs. The range of the
linear growth rates $(\mu,\nu)$ over which QCs are the global minimum of the
free energy is relatively small, but {expands} when~$Q$ is larger or $\sigma_0$
is less negative.

In conclusion, we have demonstrated that {the nonlinear resonant mechanism that operates in 2D also stabilizes} 3D icosahedral QCs as global minima {of the free energy}. This success will guide our future work in analyzing the formation of QCs in polymeric systems using realistic {dynamical density functional theory}, extending the theory from \citep{Archer2015} to three dimensions. Another avenue to explore lies in characterizing the symmetry subspaces that are retained in a QC structure using group-theoretic methods together with identifying the members of each symmetry subspace through a weakly nonlinear analysis.

\begin{acknowledgments} 

We are grateful to Ron Lifshitz, Peter Olmsted, Daniel Read and Paul Matthews for many discussions. This work was supported in part by the National Science Foundation under grant DMS-1211953 (EK).
 

 \end{acknowledgments} 
 

\end{document}